\documentclass[conference]{IEEEtran}
\IEEEoverridecommandlockouts
\usepackage{caption}
\usepackage{cite}
\usepackage{amsmath,amssymb,amsfonts,bm,amsthm,}
\usepackage{algorithmic}
\usepackage{graphicx}
\usepackage{textcomp}
\usepackage{xcolor}

\newtheorem{prop}{Proposition}

\usepackage{subfigure}
\usepackage{tcolorbox}
\usepackage{booktabs}
\usepackage{multicol}
\usepackage{multirow}
\def\BibTeX{{\rm B\kern-.05em{\sc i\kern-.025em b}\kern-.08em
    T\kern-.1667em\lower.7ex\hbox{E}\kern-.125emX}}
    
\begin{document}
\title{
Dynamic Reserves for Managing Wind Power

\thanks{
Partial funding for this  work comes from the Advanced Research Projects Agency-Energy (ARPA-E), US Department of Energy, under Award Number DE-AR0001277.}}

\author{
\IEEEauthorblockN
{Marija Ili\'{c}, IEEE Life Fellow; 
Dongwei Zhao, IEEE Member
} 
Massachusetts Institute of Technology, Cambridge, MA, ilic@mit.edu; zhaodw@mit.edu.\\
}
\maketitle
\begin{abstract}
In this paper, we suggest that reserves must be computed dynamically to account for wind power volatility. We formalize the notion of dynamic reserves in support of sequential  Day-Ahead-Market (DAM) and Real-Time-Market (RTM) clearing and discuss several ways of accounting for wind power uncertainties. Three different wind integration protocols are considered and their impact on stakeholders, and, notably, on the need for dynamic reserves is analyzed.  An optimal strategy is proposed which requires the least reserve as long as the bids are truthful and implementable with high confidence.  This can be checked by the existing market monitor and/or by a separate risk buro. Concepts are illustrated in part using publicly available information about NYISO.  
\end{abstract}


\section{Introduction}
Flexible dynamic utilization methods are needed to manage intermittent resources, such as wind power, during both normal and contingency operating conditions \cite{nyiso}.   One way of managing this is to deploy fast-responding storage, and have stand-by reserves. However, the amount of storage and reserve needed greatly depends on how well uncertainties are accounted for by utilizing wind forecast,  and/or by scheduling wind power offered by the stakeholders.   In markets, it becomes necessary to provide right incentives for flexible operations and participation by flexible technologies, such as fast-responding power plants and storage \cite{ferc2222}.   To achieve this, it is necessary to enhance today's operating and planning protocols so that the basic technical tasks (T1. scheduling supply to meet predictable demand; T2. compensating delivery losses; T3. enabling grid delivery; T4. compensating hard-to-predict deviations from schedules; and, T5. providing reserves to ensure reliable service during equipment failures \cite{galiana})  account for growing intermittent resources, and so that the right technologies are incentivized to participate in performing these tasks at value.   

Needless to say that enhancing today's deterministic operating/market clearing practices to systematically account for significant effects of intermittency represents a somewhat monumental ongoing challenge. This paper is motivated, in particular,  by our observation that both integration protocols by the ISOs/markets, and the bidding strategies by the stakeholders interactively affect each other. To support this claim, we first  in Section \ref{Sec:dynreserve} 
elaborate on the role of temporal dynamic reserves in ensuring reliability in the changing industry. In Section \ref{Sec:windpower},  we consider the impact of three different integration protocols on wind power stakeholders.   As expected, different integration protocols result in different wind power used and different profits.  We formalize temporal aspects of dynamic reserve needed to account for the effect of intermittent resources. In Section \ref{Sec:temporal}, we propose that there exists a wind power integration protocol which both gives choice to wind power plants to decide on their own sub-objectives, and requires near-minimal dynamic reserve. In Section \ref{Sec:best}, we consider protocols for reconciling stakeholders' choices and ISO's objectives. We stress that this is only the case if bidding strategies are truthful since the probability at which reserves will be used to manage uncertain wind power would be minimized.  In Section \ref{subsec:monitor}, we stress the key role of a market monitor which in coordination with  ISO's market monitor,  will provide publicly available statistics about wind power plants meeting their committed schedules. With the right data processing, it becomes possible to estimate break-even outcomes in which profits made by the stakeholders are similar to the cost of dynamic reserves.  A Dynamic Monitoring and Decision Systems (DyMoNDS) framework offers the basis for assessing trade-offs between the two \cite{dymonds}. Notably, if a Model Predictive Control (MPC) mode is enabled by the market monitor, the resulting power imbalances are minimized by a combination of  MPC-based truthful bidding and/or MPC-based stochastic decision making by the ISOs. It has been already shown that the distributed interactive MPC is scalable and it can be used by real-world systems, such as the one in Puerto Rico during extreme hard-to-predict massive equipment failures \cite{puertorico}. We propose that similar frameworks be explored by the continental ISOs. 

\section{Temporal aspects of dynamic reserve}
\label{Sec:dynreserve}

The electric power industry has been based for a long time on a deterministic worst- approach to ensuring reliability at a reasonable cost.  This has led to standards in support of functions  $T1$-$T5$. Over time,  Supervisory Control and Data Acquisition (SCADA)-enabled computer applications have been introduced for analysis assistance and decision-making needed to perform these functions. In particular,  basic   $T1$ function is used to compute feed-forward energy scheduling given the predicted system load.  Similarly,  functions $T2$ and $T3$ are performed to compensate for delivery losses and to avoid congestion during normal operations.  Also, function $T5$ ensuring secure, reliable service during the worst- $(N-1)$ or $(N-2)$ equipment outages is done in a preventive, feed-forward manner.  Additional spinning or non-spinning reserve and transmission headroom are determined to ensure that when even the worst such contingency takes place, the service to the end users remains un-interrupted during 10 minutes while contingency is still present, and during  30 minutes even after the equipment status is brought back to normal.  Today this is being done in a static preventive manner.
These five functions performed in real-time operations are not additive, and different computer applications are used in co-optimization.  In particular, 
joint co-optimization of energy and reserve resources is conducted to perform all functions except function  $T4$  which is a feedback Automatic Generation Control  (AGC)  intended to balance hard-to-predict load deviations within the scheduling time intervals so that frequency is regulated. \footnote{Notably, there exists a grey line between today's AGC Task $T4$ and the reliable service Task $T5$ in systems with intermittent resources.}

\begin{figure*}[h]
	\centering
	\hspace{-1ex}
	\subfigure[]{
	\raisebox{-1mm}{\includegraphics[width=3.1in]{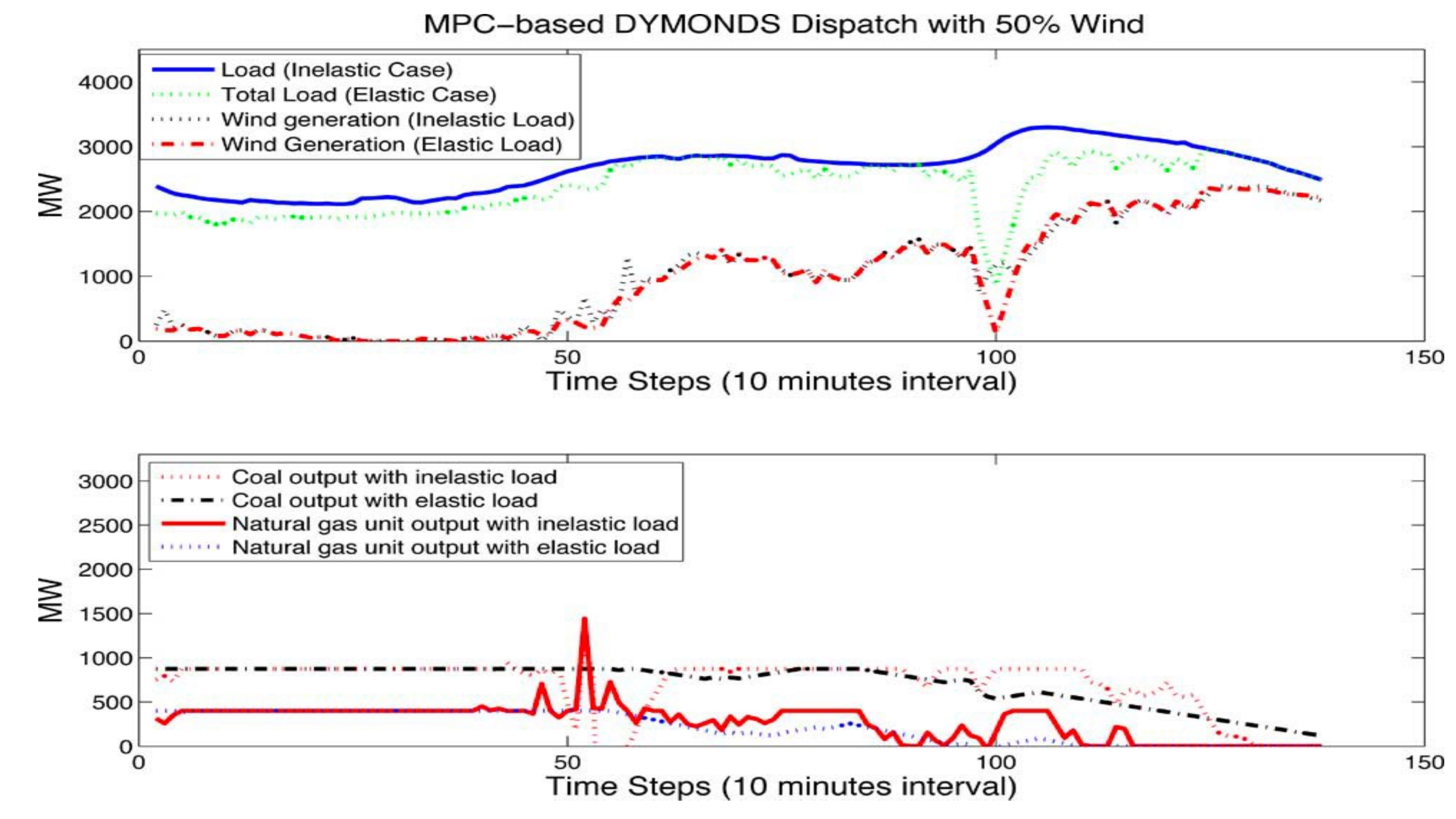}}}
	\hspace{-0ex}
	\subfigure[]{
		\raisebox{-1mm}{\includegraphics[width=3.1in,scale=.7]{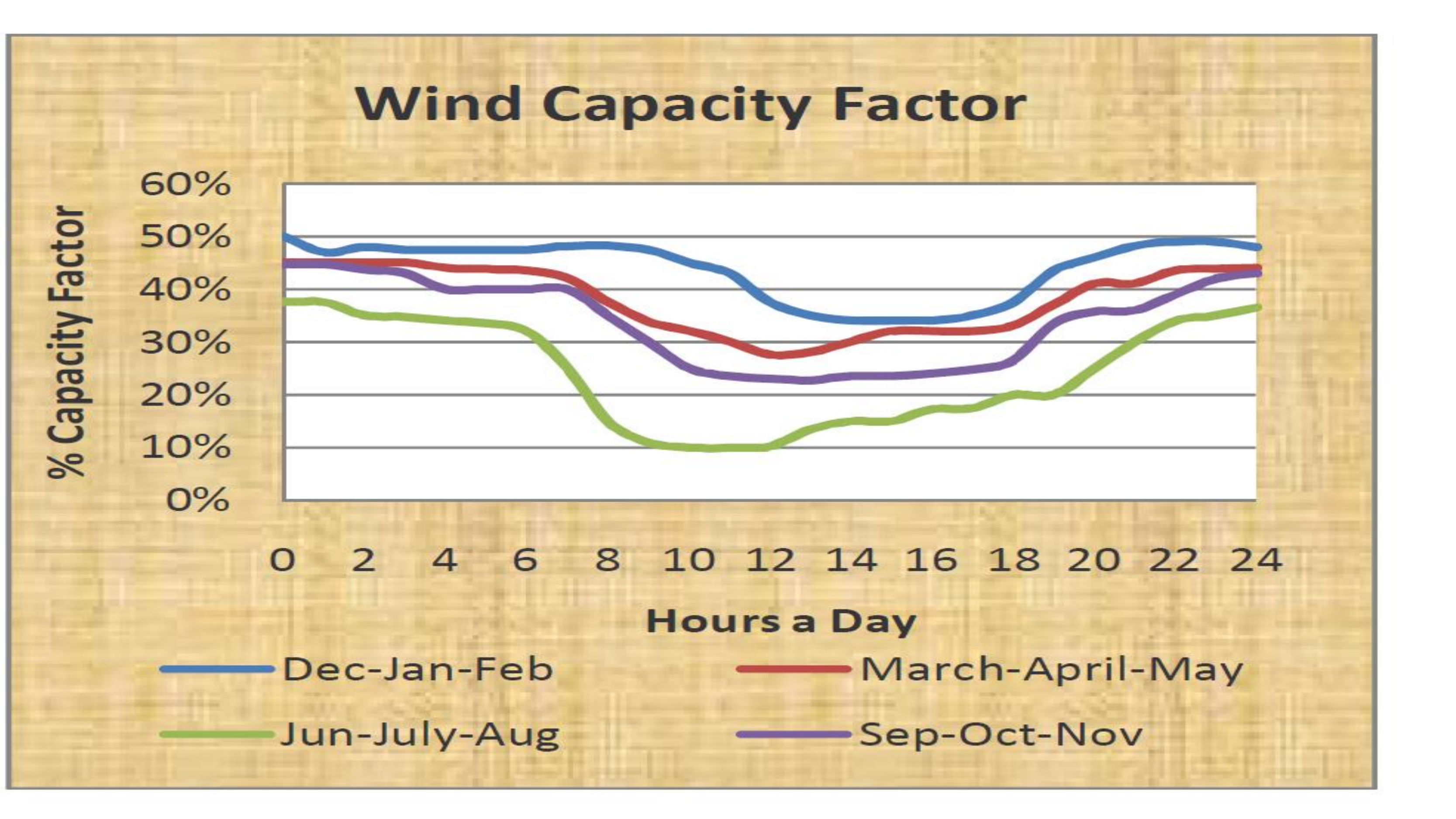}}}
\hspace{-3ex}
\caption{(a) Sudden wind gusts \cite{le}. Effect of demand response on reducing needs for fast-responding  gas power and/or storage; (b) Wind capacity factors in NYISO in different seasons \cite{nohacapfactor,windnoa}.}\label{fig:damrtmpdf}
\end{figure*}   

In this paper, we discuss challenges presented to these established operational tasks in systems with intermittent resources. To start with, it is difficult to pose the problem as a deterministic worst-case contingency without accounting for wind fluctuations shown in Figure \ref{fig:profile}. As a result, it is difficult to compute spinning- and non-spinning reserves in a preventive manner. Instead, as the wind power outputs are forecasted, (independent) System Operators ((I)SOs) are exploring ways of computing these more dynamically, and at the same time performing the other functions. For example,  the  New  York Independent System Operator (NYISO)  has proposed a notion of ``dynamic reserve" which must account for the simultaneous effect of equipment failure and sudden wind deviations \cite{nyiso}. 
This is a work in progress. UL Solutions provides as accurate forecasts as possible for assisting NYISO with this.  DAM utilizes forecasts made 72 hours ahead at hourly granularity, and 8 hours ahead of time for use in RTM at a 5-minute temporal granularity. Managing risk created by wind power fluctuations can be done in many non-unique ways. 

\section{Diverse  wind power  bidding strategies }
\label{Sec:windpower}
Wind power is fundamentally an intermittent power resource, whose control is challenging. Because of this, it is often modeled as a negative load, but its real power is much harder to predict than power consumed by conventional large-scale loads. The more fluctuating, the harder it is to balance system supply-demand and the higher need for fast-responding resources, including storage, to balance power in RTM.  Given the critical importance of knowing wind power,  to achieve this,  research is done on applying various  ML/AI  tools, in addition to many conventional data-enabled prediction methods. While high-accuracy system load prediction has been quite successful, the prediction of wind power remains a challenging problem. For example, NYISO relies on the third-party company, UI Solutions, to collect wind power forecasts as input to DAM scheduling and RTM clearing \cite{nyiso}.  

\begin{figure}[h]
	\centering
	\hspace{-2ex}
	\subfigure[]{
\raisebox{-2mm}{\includegraphics[width=1.7in,scale=.6]{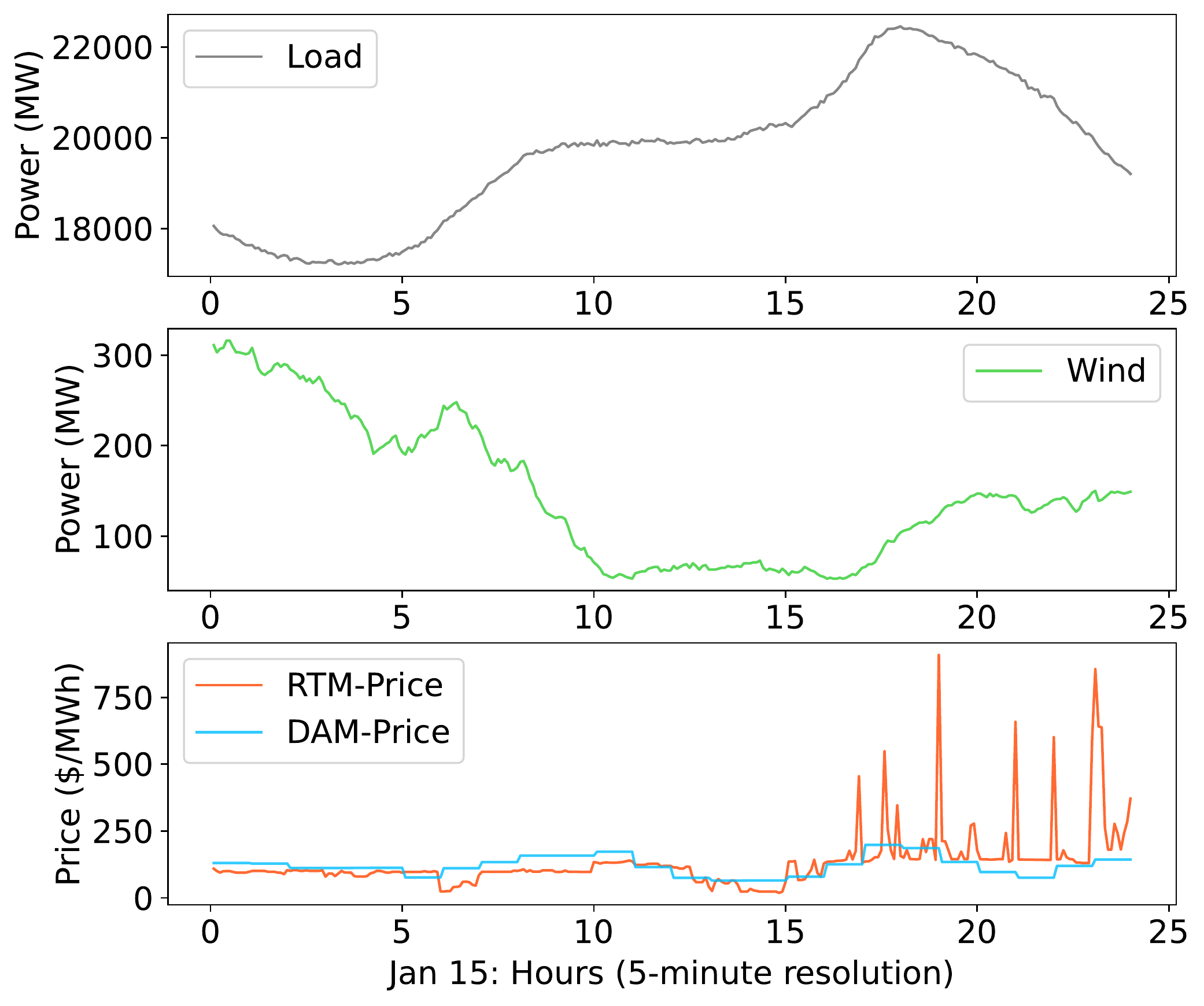}}}  
	\hspace{-2ex}
	\subfigure[]{
		\raisebox{-2mm}{\includegraphics[width=1.7in,scale =.6]{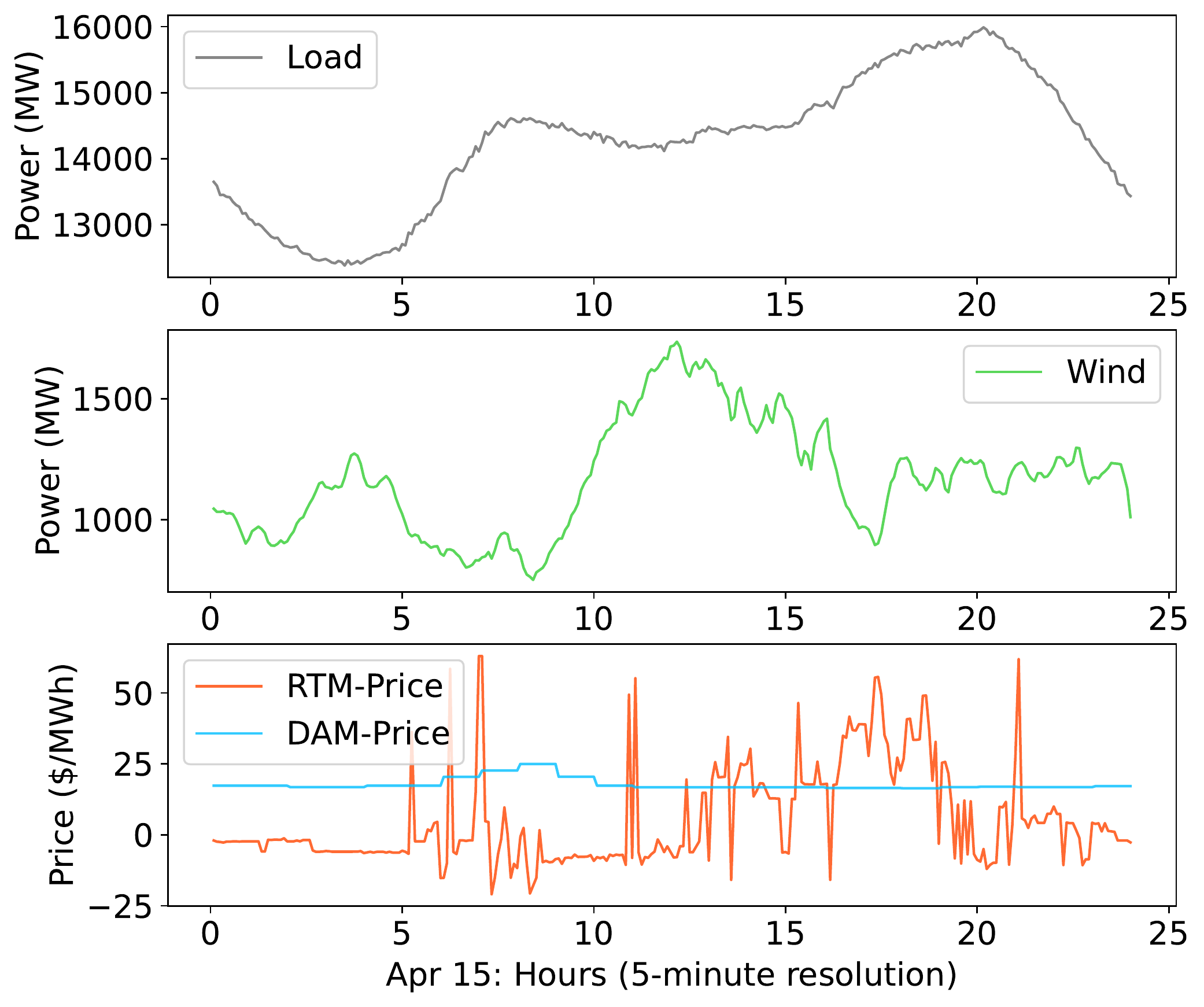}}}
	\hspace{-2ex}
	\subfigure[]{
		\raisebox{-2mm}{\includegraphics[width=1.7in,scale=.6]{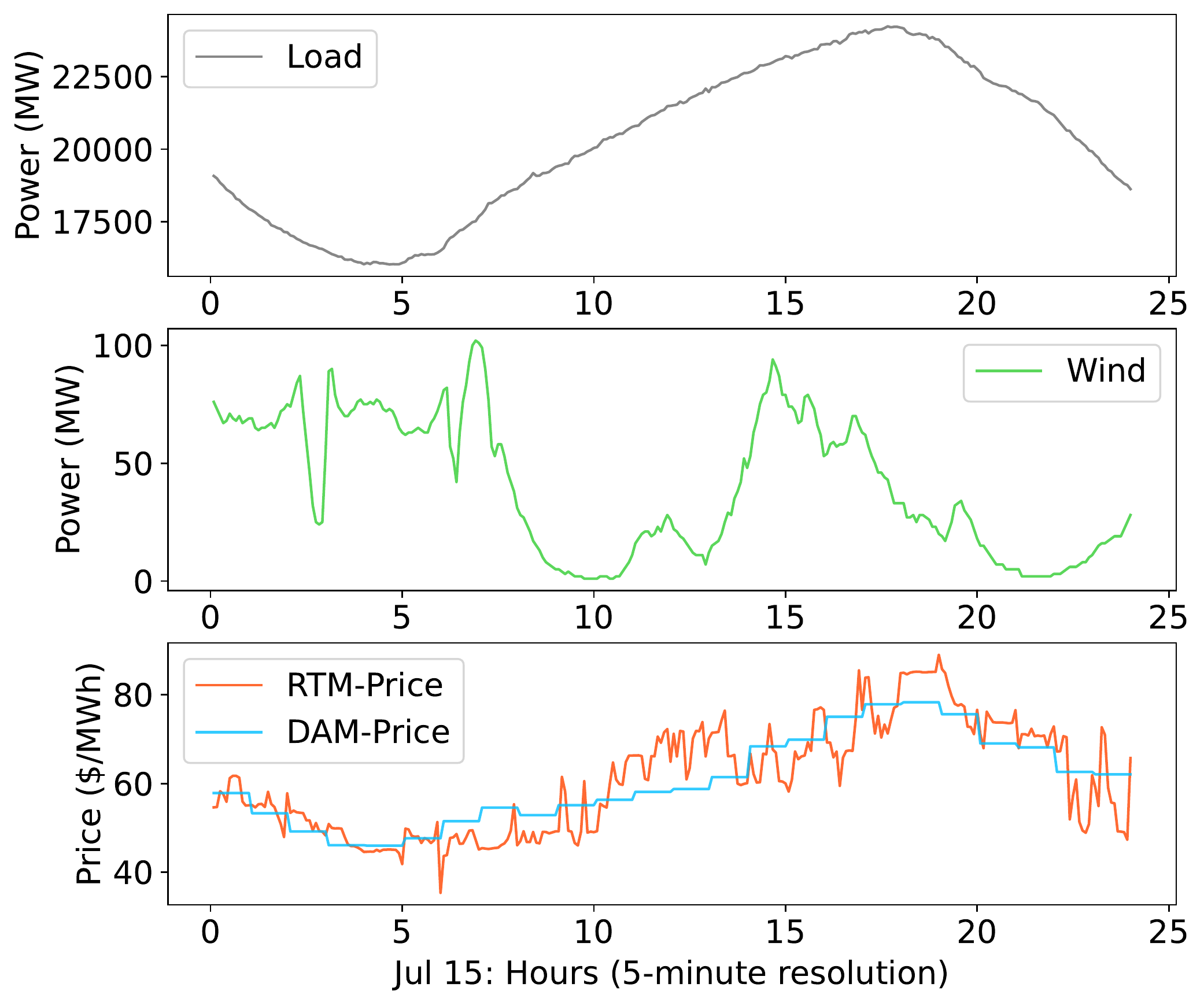}}}
			\hspace{-2ex}
	\subfigure[]{
		\raisebox{-2mm}{\includegraphics[width=1.7in,scale=.6]{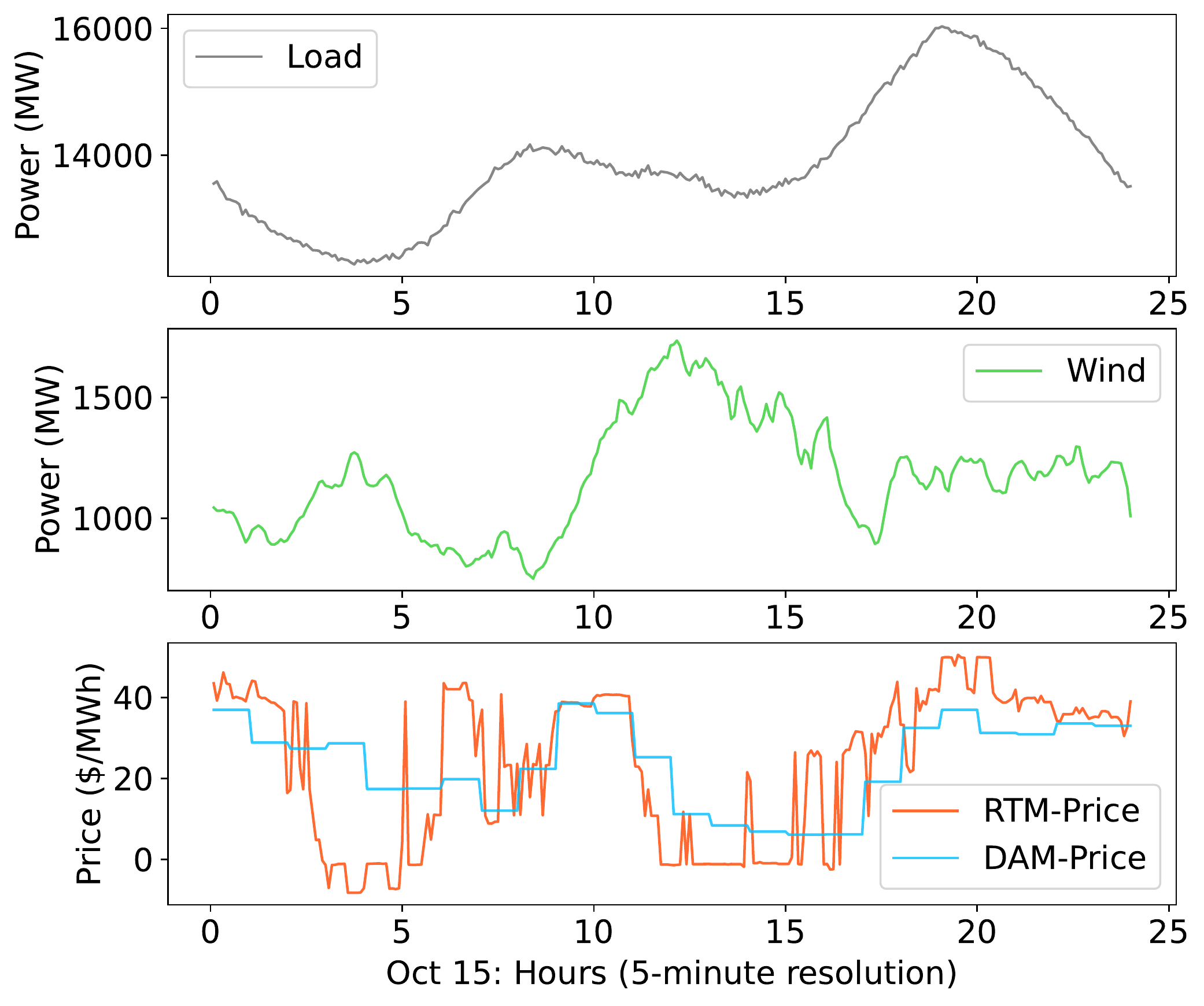}}}
	\vspace{-2mm}
	\caption{\small NYISO daily load, wind energy, RTM prices, and DAM prices on different days: (a) Jan 15 (winter); (b) Apr 15 (spring); (c)  Jul 15 (summer); (d) Oct 15 (fall) \cite{nyisowebsite}.}
	\label{fig:profile}
	\vspace{-2ex}
\end{figure}

\begin{figure}[h]
	\centering
	\hspace{-2ex}
	\subfigure[]{
	\raisebox{-2mm}{\includegraphics[width=3in]{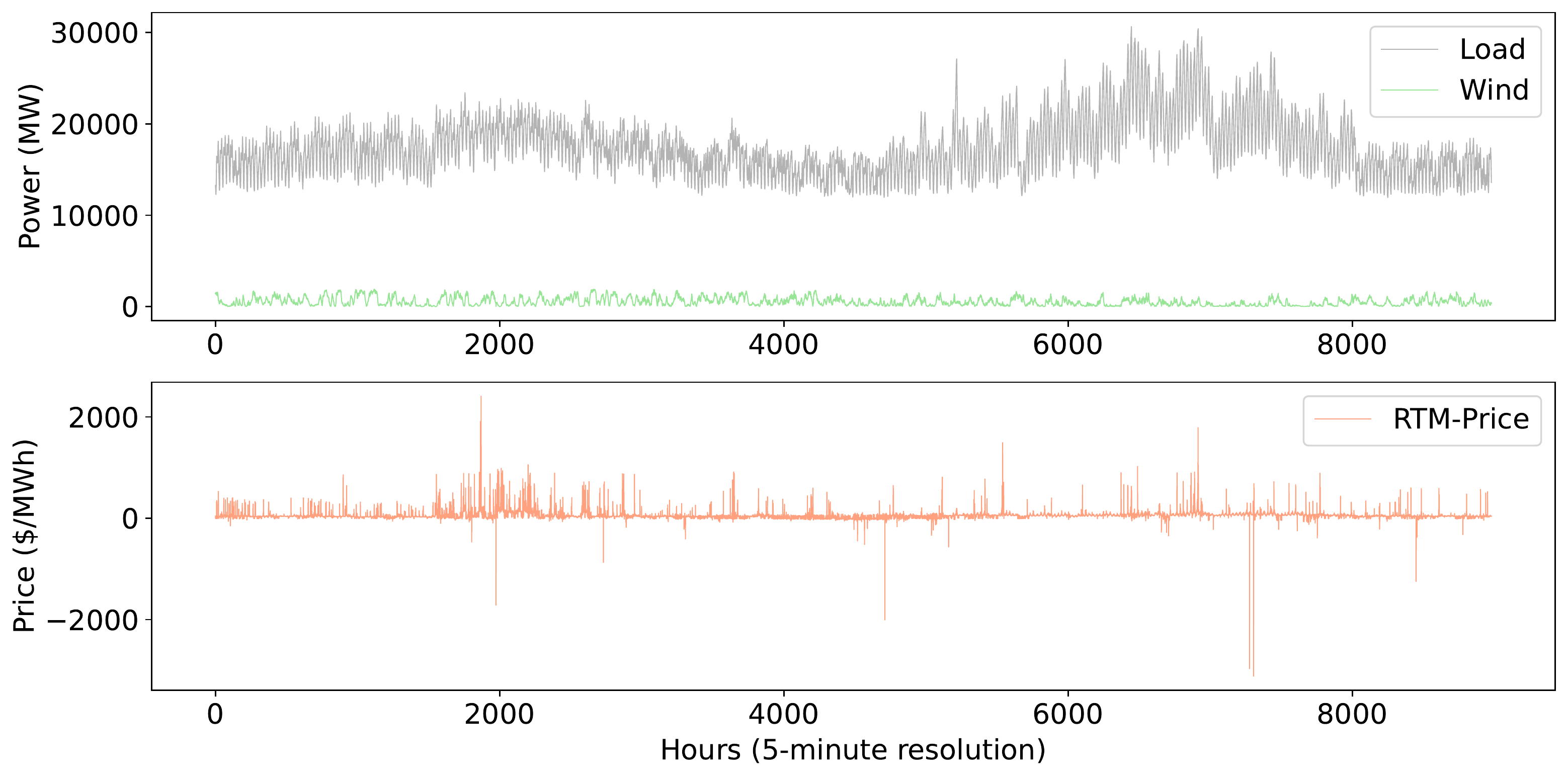}}}
	\hspace{-1ex}
	\subfigure[]{
		\raisebox{-2mm}{\includegraphics[width=3in]{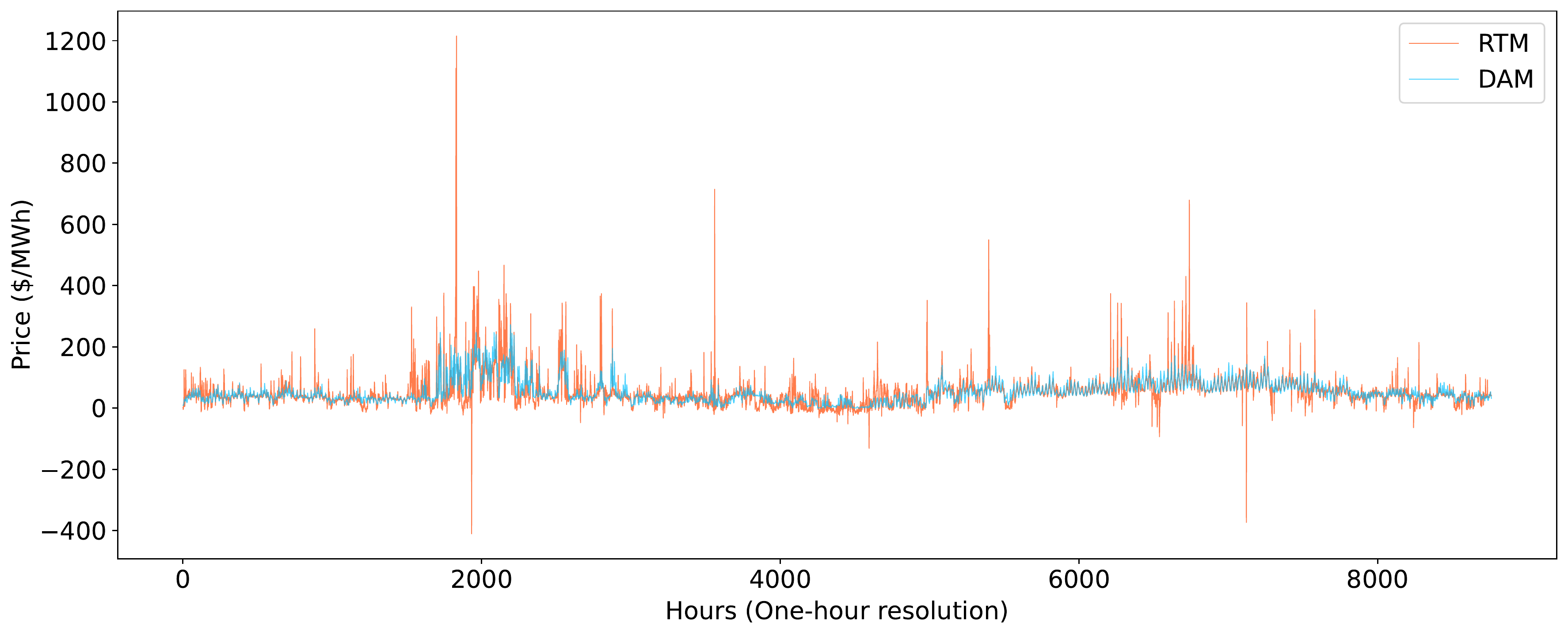}}}
		\vspace{-2ex}
\caption{\small (a) Annual load, wind, and RTM prices; (b) Annual DAM and RTM price (from Nov 2021 to Oct 2022)\cite{nyisowebsite}.}\label{fig:price-annual}
\vspace{-2ex}
\end{figure}

Shown in  Figures \ref{fig:profile}  and \ref{fig:price-annual} are the system loads and wind outputs currently seen by the NYISO,  for the entire year (from Nov 2021 to Oct 2022) and for representative seasonal days, respectively \cite{nyisowebsite}. The wind power generated is below 2 GW  and fluctuates at different rates in different seasons. We use this real data to illustrate the impact of bidding strategies on stakeholders' profits and quantities used by the system, as discussed next.

\subsection{Bidding  in uncertain technology-diverse  environment}
\label{Sec:bidding}
It is well-known that wind and solar power bidding is quite diverse and not standardized. Some stakeholders bid aggressively into DAM, while others only participate in an RTM.  Their bidding strategies are equally varying. One way or the other, unless there are rigid system constraints, such as ramp rates and congestion, wind power generally gets scheduled first, given its near-zero operating cost and also no environmental impact cost. To illustrate the impact of diverse bidding, we consider three different ways of integrating wind power into real-time operations. 

In the formulation, we denote by $\lambda^{\text{DA}}$ as the hourly DAM price. In  RTM, we denote the operation interval as  $\Delta t$, e.g., 15 minutes, and the interval set of one hour as $\mathcal{H}$. The RTM price is denoted as  $\lambda_t^{\text{RT}}$ for any interval $t \in \mathcal{H}$. For a wind power plant, we denote by $W^f$ the day-ahead forecast of wind generation and by $W_t^g$ the actual generation in real-time for any interval $t \in \mathcal{H}$. We consider three bidding strategies for wind producers.

{\bf  Scenario 1:  Wind power forecast  used in both DAM and  RTM.}
The producer bids the forecasted quantity $W^f$ in the day-ahead market, and then the actual generation deviation is settled in the real-time market. This leads to the following profit.
\begin{align}
 \lambda^{\text{DA}}\cdot  {W}^f -   \sum_{t\in \mathcal{H}} 
 \Delta t\cdot \lambda_t^{\text{RT}}\cdot ({W}^f-W_t^{g}).
\end{align}
We assume un-congested prices, wind power gets paid a uniform clearing price, much the same as all other scheduled generators. 

{\bf  Scenario 2:  Forecast wind power is scheduled in RTM only.}
This scenario is different from Scenario 1 because the wind power plant does not participate in DAM, and the overall uncertainty seen by the ISO gets reduced because wind power only participates in RTM and bids the actual quantity $W_t^g$ in the real-time market. The profit is
\begin{align}
 \sum_{t\in \mathcal{H}} \Delta t \cdot \lambda_t^{\text{RT}}\cdot W_t^g.
\end{align}

{\bf Scenario 3: Wind power provides bids into DAM.}
Wind power plants create bids by attempting to maximize their overall profit according to the following strategy
\begin{align}
\max_{0\leq q\leq W^f}~ &\lambda^{\text{DA}}\cdot  q -  \sum_{t\in \mathcal{H}} \Delta t\cdot \lambda_t^{\text{RT}} \cdot (q-W_t^{g})
\label{scenario3}
\end{align}
The objective is to maximize the revenue made in DAM  while recognizing that in RTM failure to produce will reduce the DAM profit. The producer optimizes the bidding quantity $q$ in the day-ahead market to maximize its total profit in the DAM and RTM. We have the following optimal result of decision-making in Scenario 3.
\begin{prop}
The optimal bidding quantity $q^*$ in Scenario 3 is given as follows.
\begin{itemize} 
    \item If $\lambda^{\text{DA}}\leq \Delta t \cdot \sum_{t\in \mathcal{H}} \lambda_t^{\text{RT}}$, $q^*=0$.
        \item  If $\lambda^{\text{DA}}> \Delta t \cdot \sum_{t\in \mathcal{H}} \lambda_t^{\text{RT}}$,  $q^*=W^f$.
\end{itemize}
\end{prop}
As shown in this proposition, the optimal bidding strategy is affected by both DAM and RTM prices shown in Figure \ref{fig:price-annual}. If the DAM price is no higher than the average RTM price, the producer should bid zero in the day-ahead market to take advantage of the real-time high price, otherwise, it should bid the maximum quantity in the day-ahead market to avoid the unfavorable real-time price. Note that this Scenario 3 always gives a higher profit to the producer than Scenario 1 and Scenario 2. Scenarios 1 and 2 are equivalent to the strategy $q=W^f$ and $q=0$ in Scenario 3, respectively. 
\subsection{Numerical illustration}
\begin{figure}[ht]
	\hspace{-2ex}
	\subfigure[]{
	\raisebox{-2mm}{\includegraphics[width=1.74in]{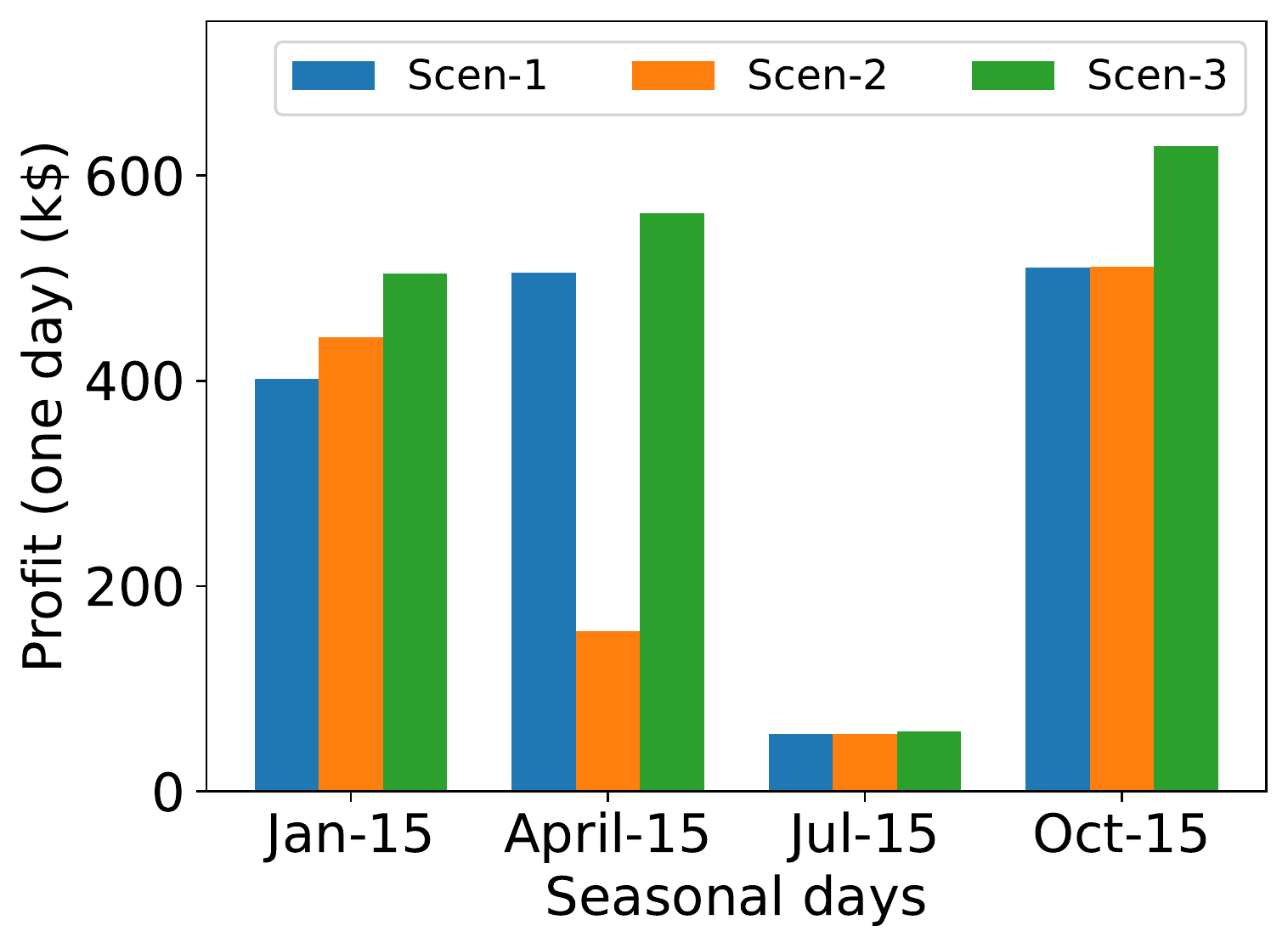}}}
	\hspace{-2ex}
	\subfigure[]{
		\raisebox{-2mm}{\includegraphics[width=1.74in]{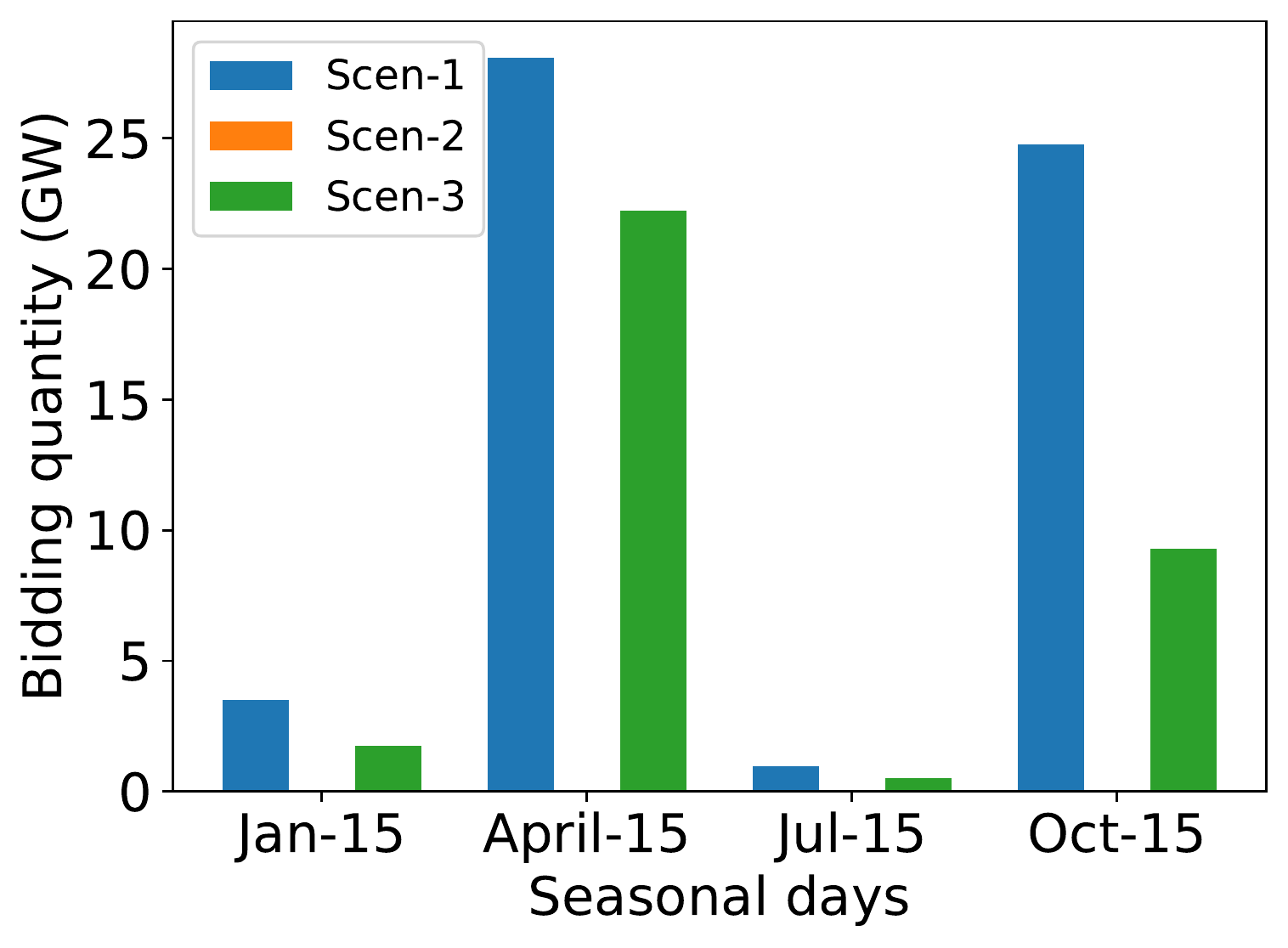}}}
	\vspace{-2ex}
\caption{(a) \small Profits on four days of different seasons; (b) \small Day-ahead bidding quantities on different seasonal days.}\label{fig:profit}
\vspace{-3ex}
\end{figure}

We examine the profits and day-ahead bidding quantities of wind producers (aggregated level) for four days of different seasons: Jan 15 (winter), Apr 15 (spring), Jul 15 (summer), and Oct 15 (fall). We consider 15-minute  windows for real-time operations and use average wind generations of one day as the forecast. Figure \ref{fig:profit}(a) shows the profits of Scenario-1 (blue bar), Scenario-2 (orange bar), and Scenario-3 (green bar) strategies on the four selected days, respectively. Figure \ref{fig:profit}(b) shows the corresponding total day-ahead bidding quantities in one day. As shown in Figure \ref{fig:profit}(a), the Scenario-3 strategy always gives the highest profit to producers, which can be over 20\% higher than the best of Scenarios -1 and -2.

It can be seen that the wind producers' profits are affected by both seasonal market prices and wind generation. On Jul 15, the profit is much lower than on other days because the wind generation amount is small (below 100 MW all day) as shown in Figure \ref{fig:profile}(c). Furthermore, the DAM and RTM prices are relatively high but not extreme (between 40 to 90\$/MWh) in summer. On Jan 15, although the wind generation amount is small (below 300 MW almost all day),  the DAM and RTM prices are very high (beyond 700\$/MWh for some time) in winter as shown in Figure \ref{fig:profile}(a), which provides high profits for producers. The day-ahead bidding quantities in  the three strategies on Jan 15 and Jul 15 are also small as in Figure \ref{fig:profit}(b). On Apr 15 and Oct 15, the DAM and RTM prices are low (below 50\$/MWh almost for all day) in spring and fall, but the wind generation amount is very high (beyond 1000 MW almost all day).  The difference between DAM prices and RTM prices will affect the profit ranking of Scdenarios-2 and -1. Especially on Apr 15, the -2 strategy will lead to a much lower profit than other strategies. The reason is that the RTM prices are overall lower than the DAM prices as shown in Figure \ref{fig:profile}(b), which can be due to a lot of hydropower in spring. This makes direct participation in the real-time market less profitable. 


\section{The impact of wind power variability  on dynamic reserve  requirements}
\label{Sec:temporal}
Based on the discussion in the above section, we suggest that wind power deviations can occur for a variety of reasons beyond weather factors typically accounted for when making the forecast.  Notably, different bidding strategies by heterogeneous wind power plants as analyzed in Section \ref{Sec:windpower} above, result in different power actually used in real time. 

\begin{figure}[h]
\includegraphics[width=3.43in,scale = .5]{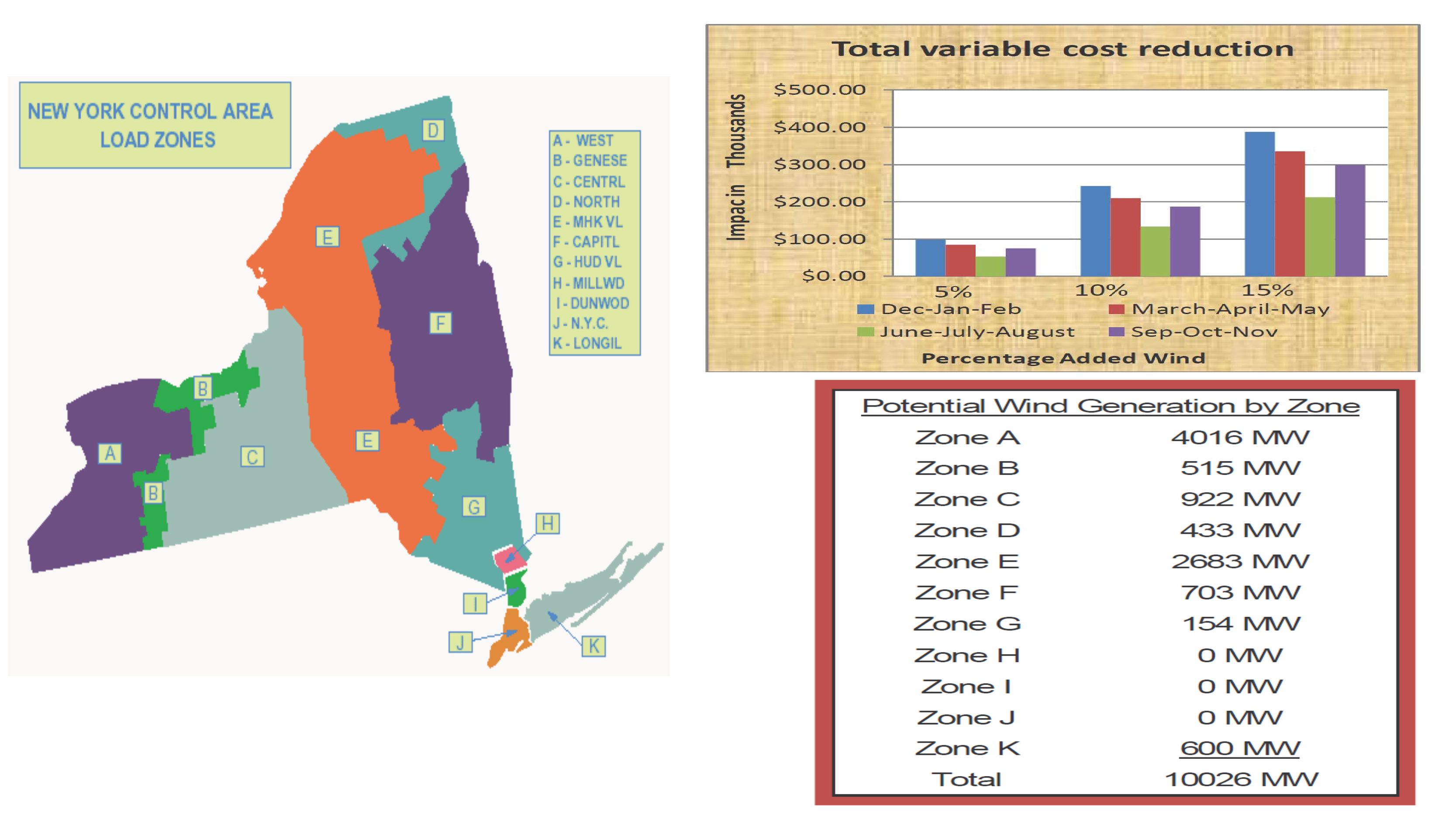}
\vspace{-0.5ex}
	\caption{\small Variable cost saving in NYISO system\cite{nohailic}.}
 \label{fig:nohailic}
	\vspace{-2ex}
\end{figure}

The uncertainties caused by deviations of scheduled or forecast wind power generation from the actual are becoming a  big concern to ISOs during contingencies \cite{nyiso}.  Because of this, it has been proposed to account for a sudden loss of wind power generation when  $(N-1)$ or $(N-2)$ contingency screening is done using well-established Security Constrained Economic Dispatch (SCED)/Security Constrained Unit Commitment (SCUC). For different wind capacities recorded in NYISO as shown in Figure \ref{fig:damrtmpdf} b), even the basic economic dispatch optimizes generation during normal conditions \cite{nohailic}. Given these typical wind capacities in NYISO \cite{windnoa} shown in Figure \ref{fig:damrtmpdf}, economic dispatch using somewhat outdated generation data in NYISO \cite{allen} results in total generation cost savings shown in Figure \ref{fig:nohailic}.  It can be seen that the economic outcomes are quite sensitive to how accurate the wind capacity factor is. This makes ISO decisions quite difficult when setting aside the reserve necessary to meet today's  $(N-1)$ reliability criteria. The ISO can best justify the cost of reserve when the capacity factor is known.  Also, it can be seen from Figure \ref{fig:damrtmpdf} (b) that the variability of capacity factors is quite significant.  This requires recomputing reserve in anticipation of wind capacity factor and the equipment failures more dynamically in DAM and RTM. A major challenge, not discussed in this paper, concerns locational aspects of deliverable reserves \cite{itroops}. A quick look at the NYISO website shows that the congestion cost dominates energy cost in zones J and K of NYISO, Figure \ref{fig:nohailic}. This is in part a result of zonal reserve requirements.  These are going to drastically change with the planned deployment of off-shore wind in Long Island. It follows that the more predictable short-and long-term wind power capacity factors are, the lower requirements for expensive congestion-caused reserve requirements become.  While beyond the objectives of this paper, such a study should be carried out as it greatly affects the resulting benefits from wind power deployment. \footnote{These economic saving estimates are obtained using homegrown GYPSIS DC Optimal Power Flow at Carnegie Mellon, for details see \cite{nohailic}. Optimizing deliverable reserves as more wind power is deployed in Long Island must be done using AC Optimal Power Flow, since major bottlenecks to deliverable reserve in NYISO  are voltage-related problems, as documented in \cite{nyserda}. We estimate that the cost of deliverable reserves obtained using AC OPF will greatly  be reduced when voltage is optimized.}

\section{Possible protocols for  reconciling  stakeholders' choice and ISO requirements}
\label{Sec:best}

Fundamentally,  there exists a break-even between the benefits of utilizing more wind power and the costs of dynamically providing reserve to ensure reliable service despite wind power deviations. This trade-off can be computed after the fact and can be used to provide incentives for operating close to this break-even schedule.  It is important to observe that if the wind power plant bids truthfully at a high probability that their bids would be physically implementable as committed and scheduled, then, as with any other type of controllable generation, the less dynamic reserve would be needed, all else being the same.  The cost of reserve is generally quite high, and, because of this, we conclude that Scenario 3 bidding strategy probably comes the closest to the break-even optimal real-time operations.  While earning more than wind power plants that follow bidding Scenarios  1 and 2,  Strategy 3 is still better when measuring the overall deliverable energy and reserve dispatch cost.  We also observe from the analysis in Section \ref{Sec:windpower} that it is not always optimal to bid the maximum forecast power either into DAM or RTM. The outcomes are highly sensitive to underlying uncertainties, and one's ability to schedule implementable wind power generation.  
Notably, having the worst-case approach to scheduling wind power for co-optimized energy and reserve is not optimal.  The frequency of deviating from expected, worst wind power is not uniform over seasons and throughout days. Taking the overall conclusions in this paper,  it follows that wind power bidding according to own sub-objectives, and having protocols for assessing how truthful their bidding is,  will lead to minimizing the cost of reliable service in systems with higher wind power generation.  
\subsection{The role of a market monitor  and/or a new Risk Buro}
\label{subsec:monitor}
To ensure truthful bidding, the market monitor can and should work closely with wind power generators to understand, audit, and certify their bidding strategies. Given very diverse risk preferences and wind generation and control technologies, market monitor should be responsible for ensuring that there is no significant un-justified gaming for making excessive profits through bidding. At present levels of wind power, it is highly unlikely that the wind power bidding can significantly affect  DAM clearing price.  
 The process of market monitors interacting directly with different stakeholders to approve bidding beyond their short-marginal costs has already begun.  Hydropower plants in New York Power Authority (NYPA) are allowed to bid higher than their short-run marginal costs because of the key role as storage.  It is only a matter of time when diverse storage will be allowed to bid according to model predictive control (MPC) strategies. Even conventional power plants should be allowed to bid in an MPC manner because they contribute to managing wind variations by scheduling their outputs according to their ramp rates which benefit them over the longer time horizons. Scheduling when the market price is low, to prepare to sell when the price is higher has been shown to be quite beneficial to supporting large deviations in renewable power and, to, at the same time maximizing their longer-term profits \cite{ferc}.  This look-ahead  MPC bidding for the best forecast possible plays effectively the role of fast gas power plant scheduling in real time and/or utilizing fast storage.  But, the clearing price with reasonable wind power generation predicted is fundamentally much lower than when the bidding protocol is to only bid short-run marginal cost and not account for uncertainties. 

It has also been proposed recently that instead of relying entirely on forecasts by a third party and on market monitor,  one could have an open  Risk Buro platform which works closely with all market participants, the ISO, stakeholders, market monitor, and companies providing forecasts.  Its main objective is to support interactive information exchange between different layers of an otherwise fractured environment. Data-enabled decision making and interactive information exchange result in market-clearing outcomes (quantities, prices)  which are further post-processed to assess how truthful stakeholders are in creating their bids, and how accurately they actually meet their schedules.  Buro creates repositories of such data, processes various correlations, and evolves into  Standard and Poor's (S\&P's) for the changing electric energy industry.  In particular,  a Risk Buro produces, in collaboration with the market monitor, indices that help classify stakeholders according to their performance.  ISOs can use this information when computing the dynamic reserve needed and the like. This DyMoNDS-based  Risk Buro represents a platform that facilitates the self-adjusting of all market participants in an MPC manner, with stakeholders bidding according to their sub-objectives and aligning with  ISO-level objectives.  
\subsection{Closing takeaways \cite{3rs}}
\begin{itemize} 
\item Needs and resources can be accurately forecasted only when producers and responsive demand provide binding self-commitments.
\item (Today) The probability of utilizing full capacity is very low and resources are, by and large, under-utilized.
\item Internalizing the risk by those creating it would make the responsibilities much better understood.
\item ``Solar (and wind) power makes lots of sense, but utilities are used to dealing with rotating equipment and this is a whole new animal." Ryan Sather, Accenture, 2009. 
\end{itemize}

\end{document}